\documentclass[aps,preprint,showpacs,showkeys]{revtex4}
\usepackage{latexsym}
\usepackage[xdvi]{graphicx}
\usepackage{amssymb,amsmath}
\newcommand{\bsl}[1]{\begin{slide}{#1}}
\newcommand{\esl}{\end{slide}}
\newcommand{\be}{\begin{equation}}
\newcommand{\ee}{\end{equation}}
\newcommand{\ben}{\begin{enumerate}}
\newcommand{\een}{\end{enumerate}}
\newcommand{\bit}{\begin{itemize}}
\newcommand{\eit}{\end{itemize}}
\newcommand{\been}{\begin{displaymath}}
\newcommand{\eeen}{\end{displaymath}}
\newcommand{\ba}{\left[\begin{array}}
\newcommand{\ea}{\end{array}\right]}
\newcommand{\bac}{\begin{array}}
\newcommand{\eac}{\end{array}}
\newcommand{\bc}{\begin{center}}
\newcommand{\ec}{\end{center}}
\newcommand{\bea}{\begin{eqnarray}}
\newcommand{\eea}{\end{eqnarray}}
\newcommand{\bean}{\begin{eqnarray*}}
\newcommand{\eean}{\end{eqnarray*}}
\newcommand{\bqu}{\begin{quote}\begin{it}}
\newcommand{\equ}{\end{it}\end{quote}}

\begin{document}
\title{Lattice distortion in hcp rare gas solids}
\author{A. Grechnev}
\affiliation{B. Verkin Institute for Low Temperature Physics and
Engineering, National Academy of Sciences, 47 Lenin Ave., 61103
Kharkov, Ukraine}
\author{S. M. Tretyak}
\affiliation{B. Verkin Institute for Low Temperature Physics and
Engineering, National Academy of Sciences, 47 Lenin Ave., 61103
Kharkov, Ukraine}
\author{Yu. A. Freiman}
\affiliation{B. Verkin Institute for Low Temperature Physics and
Engineering, National Academy of Sciences, 47 Lenin Ave., 61103
Kharkov, Ukraine}

\begin{abstract}
The lattice distortion parameter $\delta \equiv c/a-\sqrt{8/3}$
has been calculated as a function of molar volume
for the hcp phases of He, Ar, Kr and Xe.
Results from both semi-empirical potentials and density functional theory 
are presented. Our study shows that 
$\delta$ is negative for helium in the entire pressure range. For Ar, Kr and Xe,
however, $\delta$ changes sign from negative to positive as the pressure increases,
growing rapidly in magnitude at higher pressures.
\end{abstract}

\keywords{rare-gas solids, lattice distortion, two and tree-body interatomic forces, DFT calculations\\
E-mail: shrike4625@yahoo.com}

\maketitle

\section{introduction}

Rare gases (He, Ne, Ar, Kr, Xe and Rn) crystallize in the most
closely packed structures: the face-centered cubic (fcc) and the
hexagonal close-packed (hcp). Solid helium has hcp structure at
low pressures and it has been demonstrated experimentally that
it stays in this structure up to at least 57 GPa \cite{loubeyre93prl71:2272}, apart from
narrow bcc and fcc areas
at the melting line. On the other hand, heavier rare gas solids (RGS's)
have fcc structure at low pressures. Under pressure, Ar \cite{errandonea06prb73:092106}, Kr 
 \cite{errandonea02prb65:214110} and Xe \cite{jephcoat87prl59:2670} have been
found experimentally to change into the hcp structure, 
while neon stays fcc \cite{dewaele08prb77:094106}.
It has been shown in our previous paper \cite{freiman09prb80:094112} that 
the hcp structure of solid He is stabilized by strong zero-point
vibrations.

The $c/a$ ratio of the hcp RGS's is very close to the ideal value
$\sqrt{8/3} \approx 1.633$. The deviation of the $c/a$ ratio from
the ideal value is described by the lattice distortion parameter
$\delta \equiv c/a - \sqrt{8/3}$. If $\delta >0$, the lattice is
elongated along the $z$ axis, if $\delta <0$, then it is contracted 
along the $z$ axis. The point $\delta =0$ ($c/a=\sqrt{8/3}$) corresponds to
the packing of hard spheres. In the case of crystals with $c/a$
different from $\sqrt{8/3}$, the lattice has exactly the same
symmetry as the ideal hcp lattice, i.e. the space groups of these
crystals are identical for all values of $c/a$.
While most physical properties of the hcp solids are not very sensitive
to the $c/a$ ratio (for a given volume),
$\delta$  determines the second-order contributions to the crystal
field \cite{kranendonk:book}, which is zero for the ideal hcp lattice.
It is interesting to note that $\delta$
is negative for all metallic elemental hcp solids except Zn and Cd.
Our previous calculations \cite{freiman09prb80:094112} show that for hcp He $\delta$
is negative in a wide range of pressures.

While there have been numerous experimental attempts to
determine $c/a$ in helium (see the references and discussion in Ref.
\cite{freiman09prb80:094112}), the results for heavier RGS's are very scarce.
The x-ray diffraction studies of Xe by Caldwell \emph{et. al.} \cite{caldwell97sci277:930}
and Jephcoat \emph{et. al.} \cite{jephcoat87prl59:2670}
find $\delta$ to be positive under pressure (although the experimental precision was
not sufficient to prove this fact unambiguously).
On the theoretical side, there have been several papers devoted
to the calculation of the $c/a$ ratio for Ar, Kr and Xe, however the results are
also somewhat controversial. Schwerdtfeger \emph{et. al.}
\cite{schwerdtfeger06prb73:064112} investigated rare gas solids with
a number of different extended van der Waals pair potentials and found $\delta < 0$ for 
hcp He, Ne, Ar and Kr for all potentials at zero pressure. 
Caldwell \emph{et. al.} \cite{caldwell97sci277:930} and
Yao and Tse \cite{yao07prb75:134104} used density functional
theory (DFT) and found $\delta > 0$ for Xe, while 
Cohen \emph{et. al.} \cite{cohen97prb56:8575} reported negative $\delta$ for Xe, also
within the framework of DFT. It seems that we still know very little about the $c/a$ ratio
in hcp rare gas solids in the age when all structural properties of most
elemental solids (including high-pressure phases) are firmly established from
both theory and experiment.

In an attempt to clarify this issue, in the present paper we calculate
the lattice distortion parameter $\delta$ for hcp Ar, Kr and Xe as a function of
volume. In addition, we also present
results for He, which are the extension of our previous calculations \cite{freiman09prb80:094112}.
Unfortunately, we are aware of no single theoretical method which gives
an accurate $\delta$ for RGS's
both at low pressures and at very high pressures (of the order of the metallization point).
The reason for this is the different nature of the chemical binding of RGS's
(and other molecular crystals)
for different pressures. At low pressures, an RGS is a crystal with pure molecular
binding, held together by the van der Waals forces. At higher pressures,
the binding becomes more covalent in character, and finally metallic above the metallization
point. That is why we use two methods in the present paper:
semi-empirical potentials
\cite{loubeyre87prl58:1857,loubeyre88prb37:5432,freiman07fnt33:719}
and the density functional theory
(DFT) \cite{hohenberg64pr136:B864,kohn65pr140:A1133}. 

The semi-empirical (SE) potentials (with pair and triple forces included)
work very well at low pressures, while for the higher pressures
higher order $n$-body forces are important. In particular, the SE
potentials become useless in the metallic phase (where the sum over $n$-body
terms converges extremely slowly) or near the metallization point.
Density functional theory is formally an exact theory, but for
practical calculations one always needs some kind of a model expression 
for the electronic exchange-correlation energy as a functional of the electronic
density. The most widely used approximations are the 
local density (LDA) and generalized gradient (GGA) approximations. 
They are routinely used nowadays to calculate various electronic
and structural properties of all kinds of solids, including RGS's 
\cite{caldwell97sci277:930,yao07prb75:134104,dewhurst02prl88:075504}. These approximations are
expected to be rather accurate at high pressures around the metallization
point (as LDA and GGA are in general most suitable for metallic and covalent solids),
but fail at low pressures due to the poor description of the 
van der Waals (vdW) interaction \cite{tkachenko08prb78:045116}. In other words,
the two methods employed by us complement each other: one can use the semi-empirical
approach for low pressures and DFT for higher pressures.

\section{\label{sec:method}Method}
\subsection{Semi-empirical potentials}
In our semi-empirical calculations we include 
pair ($U_2$) and triple ($U_3$) interatomic
forces, therefore the expression for the total energy is  $U_{tot} = U_2+U_3$.
The $n$-body interactions with $n>3$ are not included, which makes the method
accurate only at low pressures, as we discuss below in section \ref{sec:results}.
We use the Aziz expression \cite{aziz79jcp70:4330} for the pair potentials with parameters
from Ref. \cite{freiman07fnt33:719}.
The three-body potential is taken as a sum of the long-range
Axilrod-Teller dispersive interaction and the short-range three-body
exchange interaction in the
Slater-Kirkwood form
\cite{loubeyre87prl58:1857,loubeyre88prb37:5432,freiman07fnt33:719}. We restrict
ourselves to $T = 0$ K. The zero-point energy is treated approximately
within the Einstein model. For the heavier RGS's this
approximation is valid for the whole pressure range, but for helium we exclude
the small pressure range ($\sim 0.1$ GPa) where
quantum-crystal effects play a decisive role. 
The exactly same model has been employed previously to calculate the 
equations of state for all the RGS's \cite{freiman07fnt33:719},
and the results are in excellent agreement with experiment.

Assuming that $\delta$ is small, we expand the ground state energy
$E_{gs}$ to the second order in $\delta$:
\begin{equation}
E_{gs}(\delta) = b_0 + b_1 \delta + b_2 {\delta}^2, 
\end{equation}
where the coefficients $b_0$, $b_1$ and $b_2$ depend on
the molar volume and also on the parameters of the interatomic potential.
The minimum of $E_{gs}$ is reached for
$\delta=-b_1/(2b_2)$.
Thus, in order to find the $\delta(V)$ dependence one has to calculate
the quantities $b_1(V)$ and $b_2(V)$.
The first shell of neighbors gives the main positive contribution
to $b_2$, and the sum over spheres of neighbors converges rather rapidly.
However, despite the relatively 
short-range character of the interatomic interactions,
one has to include a large number of neighbor shells 
in order to calculate $b_1$ accurately. The
reason is that the contributions of the first two shells
are exactly equal to zero, while the contributions from
more distant shells decrease rather slowly and tend to 
alternate in signs.
This is also the reason why $\delta$ is small,
namely $\left|\delta \right| \sim 10^{-4} \div 10^{-3}$ 
at low pressures. Note that the contributions
to $b_1$ from nearest and next nearest neighbors vanish for
different reasons. The
contribution from the nearest neighbors in the $xy$ plane
exactly cancels the contribution from the nearest neighbors below and
above the $xy$ plane due
to the equality of all the nearest neighbor distances
for the ideal $c/a=\sqrt{8/3}$.
Regarding the second shell, each of the six atoms in it
gives zero contribution individually. 
The contribution to $b_1$ from the third shell (which contains only
two atoms at the distance $R = a\sqrt{8/3}$ from the central
atom) leads to $b_1 > 0$ resulting in $\delta < 0$. 
Additionally, 18 neighbors of the fourth
shell at the distance $R = a\sqrt{3}$
makes a negative contribution to $b_1$ which overweights the
positive contribution from the third shell. As a result, the total contribution
to $b_1$ from the third and fourth shells is negative and leads to $\delta > 0$.
The contributions from further shells will in most cases lead to $\delta<0$.
To get a reliable result, we take into account 
50 shells of neighbors in all our calculations.

\subsection{First principles calculations}
For our DFT calculations we use the all-electron full-potential linear muffin-tin orbital
(FP-LMTO) code RSPt \cite{wills:fp-lmto,rspt_homepage}. This code is especially reliable for
high-pressure calculations, as it uses the "soft-core approach", i.e. core
electrons energies and wave functions are recalculated for each DFT iteration.
This means that the core levels in a crystal, especially under pressure,
are shifted from the atomic positions, giving more accurate 
total energy values. The generalized gradient approximation
(GGA) in the form of Perdew-Burke-Ernzerhof (PBE) \cite{perdew96prl77:3865} has been used.
All our calculations have been done for zero temperature neglecting the zero-point vibrations. 

The equilibrium $c/a$ ratio has been found for each molar volume $V$ by minimizing
the total energy $E(c/a)$ for fixed $V$. This function was calculated for
a number of $c/a$ points around the minimum and interpolated with a cubic spline.
The equilibrium $c/a$ and the energy $E(V)$ for the equilibrium $c/a$ were then
obtained from this spline. The pressure $p(V)$ (for the equation of state) was 
obtained from the calculated $E(V)$ points, again using a cubic spline for the
numerical differentiation.
We took special care in achieving high accuracy of our calculations, as the
typical energy differences for He, between, say $c/a=1.633$ and $1.634$ are of the order 
of $10^{-6}$ to $10^{-9}$ Ry, depending on $V$ (somewhat larger for the heavier elements).
In particular, the Brillouin zone integration has been performed by the tetrahedron method
with 368 $k$-points in the irreducible part of the Brillouin zone (3375 
$k$-points in the whole Brillouin zone). The self consistent cycle was
converged to the total energy accuracy of at least $10^{-9}$ Ry.
We have used two kinetic energy tails with energies +0.2 Ry and -0.2 Ry respectively in
our basis set; and the 32x32x32 mesh for the Fast Fourier Transform in the real space.
For Kr and Xe, a second energy set with two kinetic energy tails (-0.8 Ry and -1.5 Ry)
has been used to describe $3d$ and $4d$ states, respectively.
Our tests show that the results are basically insensitive to the particular choice of
the kinetic energy tails and they are well converged with the number of $k$-points
and other parameters of the calculation. We also applied the local density approximation (LDA)
to helium and found out that the LDA results (not shown) are very close to our GGA results.
In fact, we managed to achieve
the numerical accuracy of at least $3 \times 10^{-4}$ in determining $\delta$
(this does not include the possible systematic errors of GGA and the FP-LMTO method).

It is important to note that DFT maps the many-body problem to an effective 
one-particle Scr\"odinger-like Kohn-Sham equation \cite{kohn65pr140:A1133}.
The periodic solids are treated in $k$-space using Bloch theorem.
Compared to the real-space methods (like the semi-empirical potentials) it gives
DFT an advantage of automatically including all shells of neighbors and
all $n$-body interactions in its expression for the total energy. The main
disadvantage of DFT within LDA or GGA for rare gas solids is, as we already mentioned,
the poor description of the van der Waals interaction. Therefore, while this method is
ideal for the metallic phase, and also suitable for the insulating phase in the vicinity of the
metallization point, it does not give reliable results at low pressures, where an
RGS behaves as a pure molecular crystal and the vdW interaction is important.
\section{\label{sec:results}Results and discussion}
\begin{table}
\begin{tabular}{|c|c|c|} \hline
 \hspace*{1.2cm} & \hspace*{2mm} $V_0$ (cm$^3$/mol) \hspace*{2mm} & 
 \hspace*{1mm} $V_{MET}$ (cm$^3$/mol) \hspace*{1mm}\\ \hline 
 He & 21.0 & 0.228 \\ \hline
 Ar & 22.56 & 4.42 \\ \hline
 Kr & 27.10 & 6.48 \\ \hline
 Xe & 34.74 & 10.9 \\ \hline
\end{tabular}
\caption{\label{t:v0m} The experimental equilibrium molar volume $V_0$  ( from Ref. 
\cite{swenson:rgs} , Ch. 13,  p. 825, Table 1)
 and the theoretical
metallization point $V_{MET}$ (from our DFT-GGA calculations) for rare gas solids.}
\end{table}
In this section we present the results of our calculation. For reader's convenience,
in Table \ref{t:v0m} we list the experimental equilibrium molar volumes $V_0$ of rare gas solids
(from Ref. \cite{swenson:rgs}). Note that $V_0$ for helium is high due to the zero-point oscillations.
In the same table, the theoretical metallization
volumes $V_{MET}$ from our calculations are presented. The metallization volume
is typically somewhat overestimated by LDA/GGA
(by about 20\% for helium \cite{khairallah08prl101:106407}),
thus the metallization pressure is underestimated. This is a direct
consequence of LDA/GGA underestimating the band gap in insulators.

\begin{figure}
\includegraphics[scale=0.3]{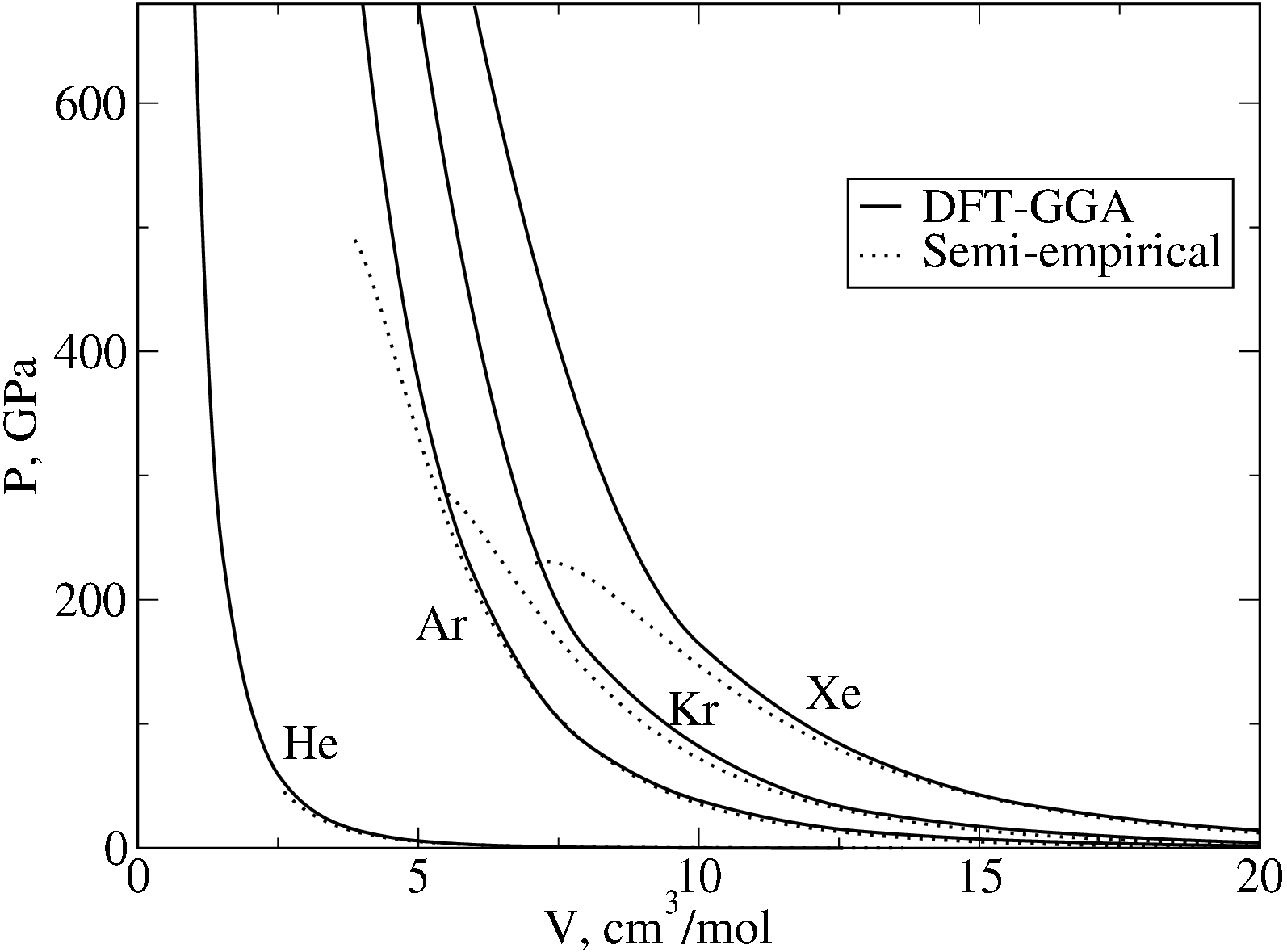}
\caption{\label{f:eqs} The calculated equation of state for He, Ar, Kr and Xe.
}
\end{figure}

The calculated equations of state for He, Ar, Kr and Xe are presented in Fig. \ref{f:eqs}, again
for the reader's convenience. The $p(V)$ curves for the equilibrium $c/a$ virtually
coincide with the $p(V)$ curves for the ideal $c/a$ (not shown). One can
see the breakdown of the semi-empirical picture at higher pressures due to the lack of the $n>3$ terms
in the $n$-body expansion.
\subsection{Helium}

\begin{figure}
\includegraphics[scale=0.3]{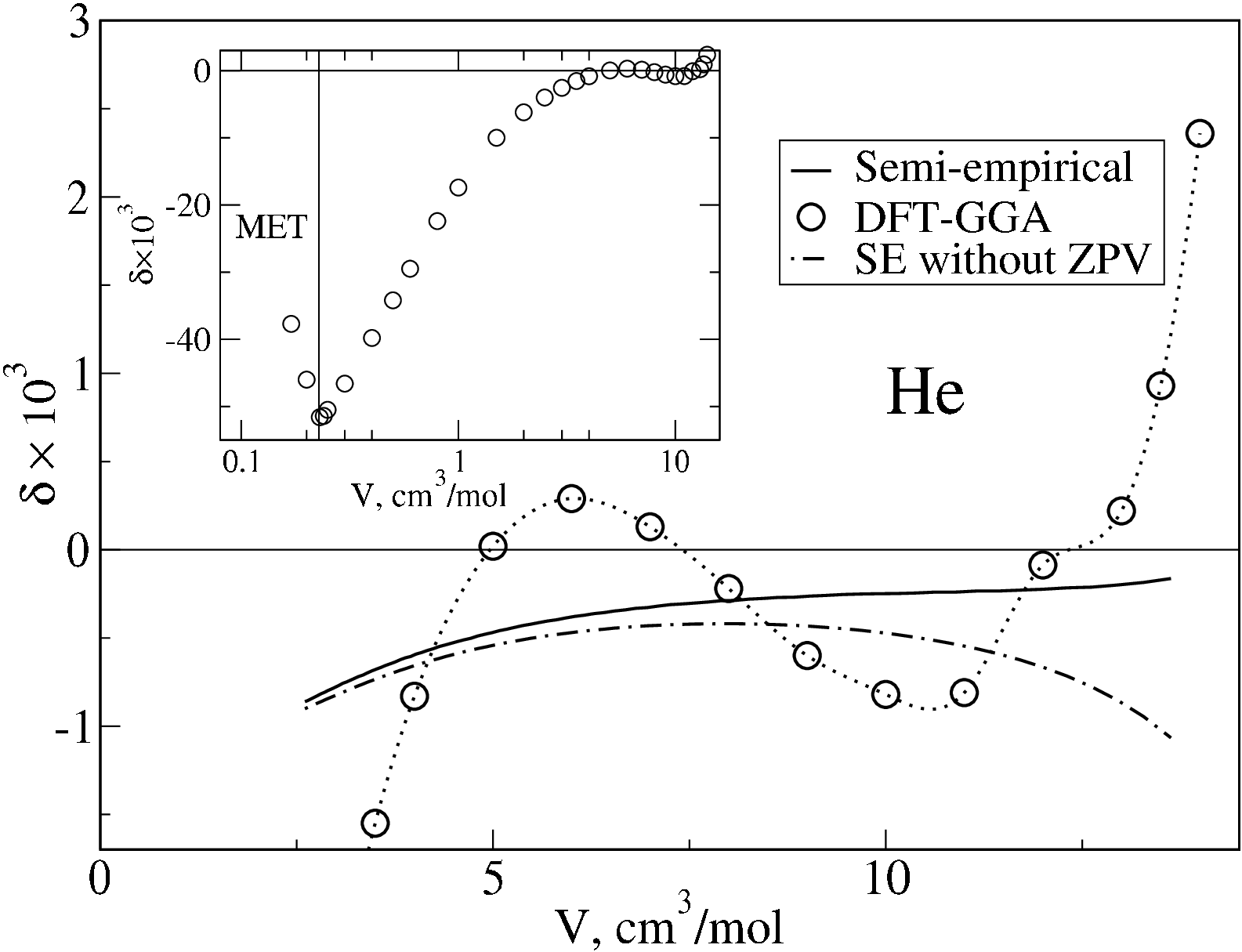}
\caption{\label{f:he}The lattice distortion parameter $\delta$ for hcp helium as a function of molar volume $V$ from
DFT-GGA, semi-empirical (SE) potentials, and the semi-empirical results without zero-point vibrations
(ZPV). The circles are calculated DFT-GGA points, while the smooth dotted curve is a cubic spline.
 The inset shows the DFT-GGA data for small volumes. 
}
\end{figure}

\begin{figure}
\includegraphics[scale=0.3]{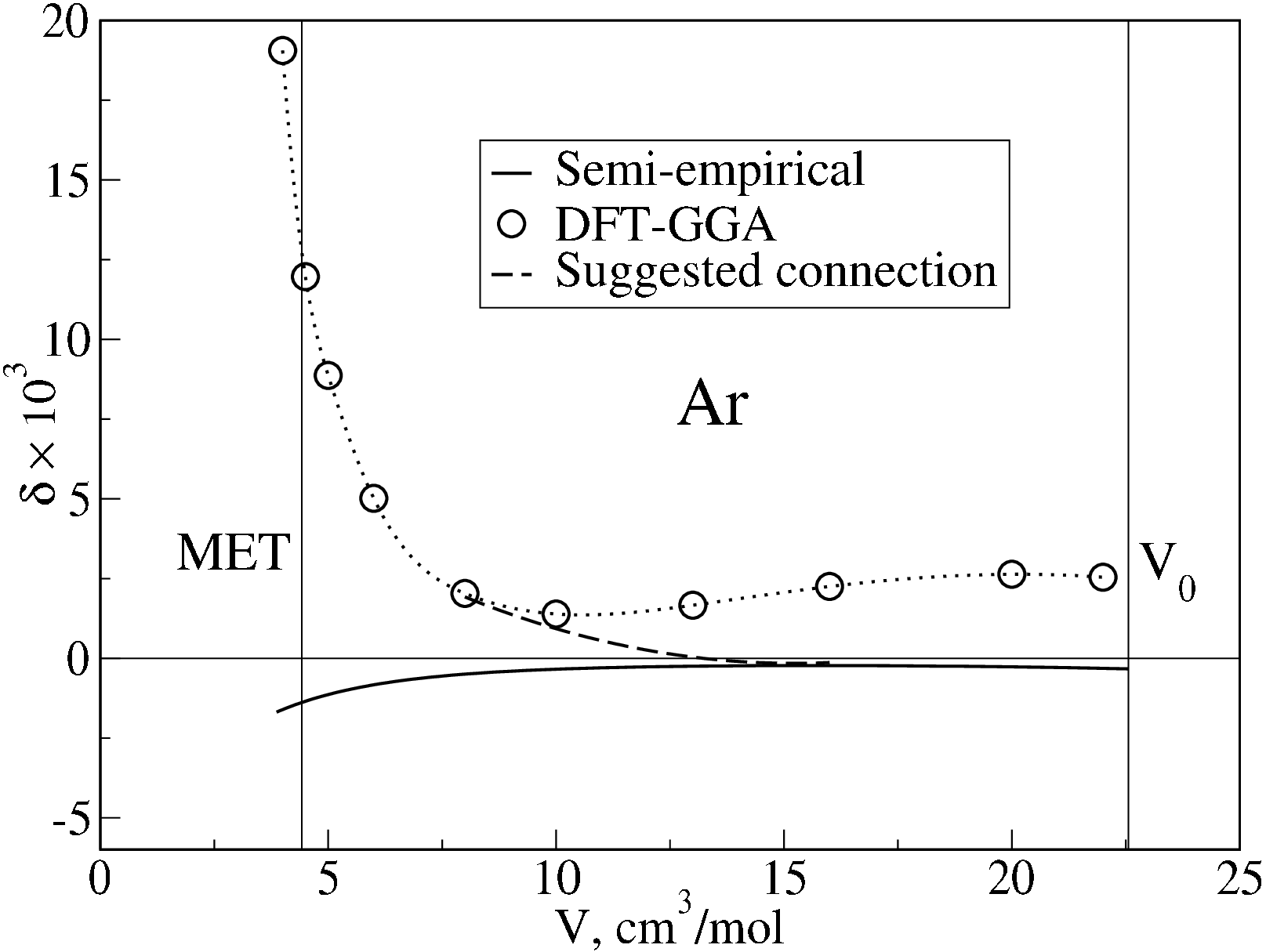}
\caption{\label{f:ar}The lattice distortion parameter $\delta$ for hcp argon as a function of volume $V$ from
DFT-GGA and the semi-empirical (SE) potentials. The circles are calculated DFT-GGA points, while the
smooth dotted curve is a cubic spline.
}
\end{figure}

\begin{figure}
\includegraphics[scale=0.3]{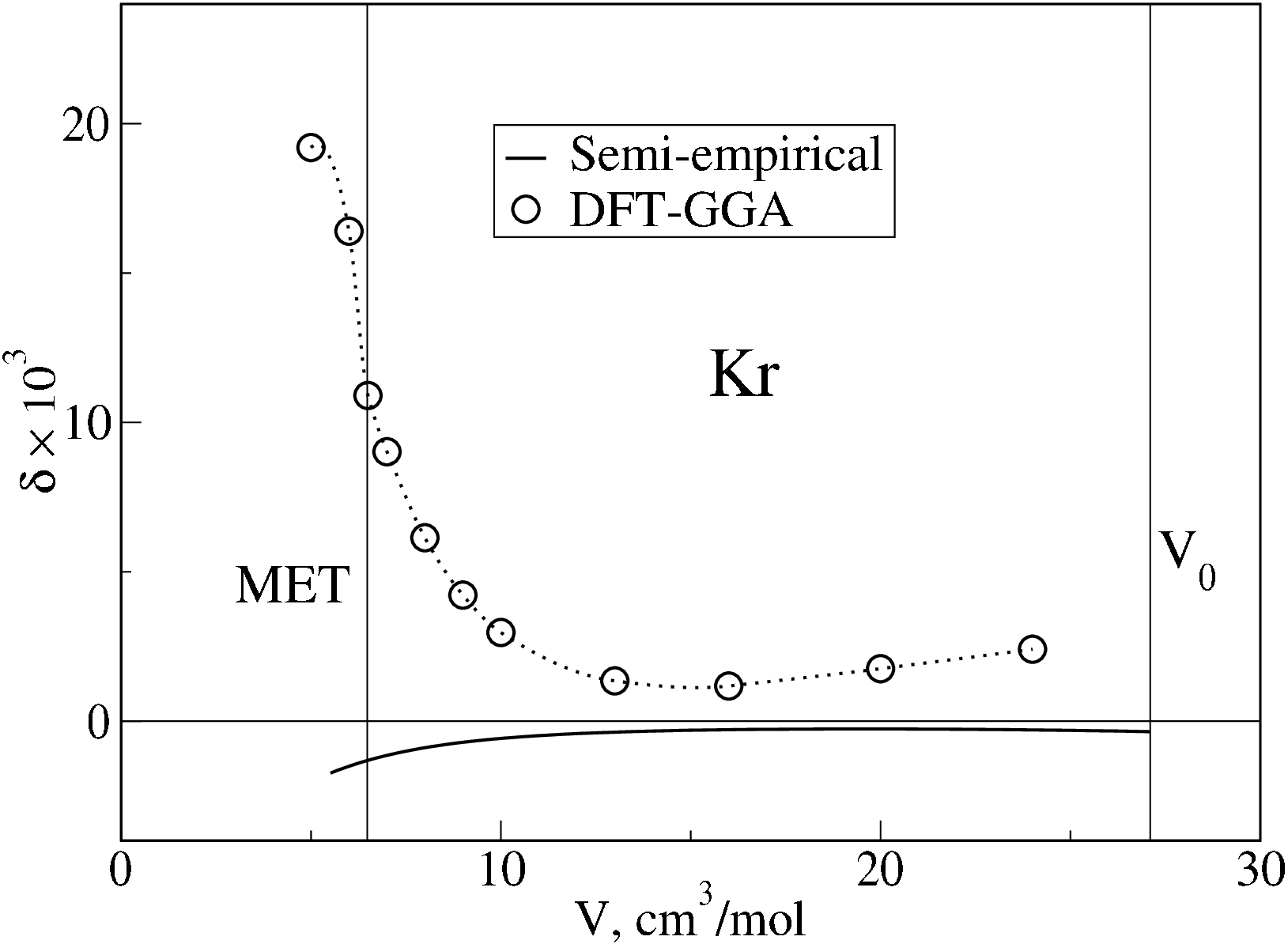}
\caption{\label{f:kr}The lattice distortion parameter $\delta$ for hcp krypton as a function of volume $V$ from
DFT-GGA and the semi-empirical (SE) potentials. The circles are calculated DFT-GGA points, while the
smooth dotted curve is a cubic spline.
}
\end{figure}

\begin{figure}
\includegraphics[scale=0.3]{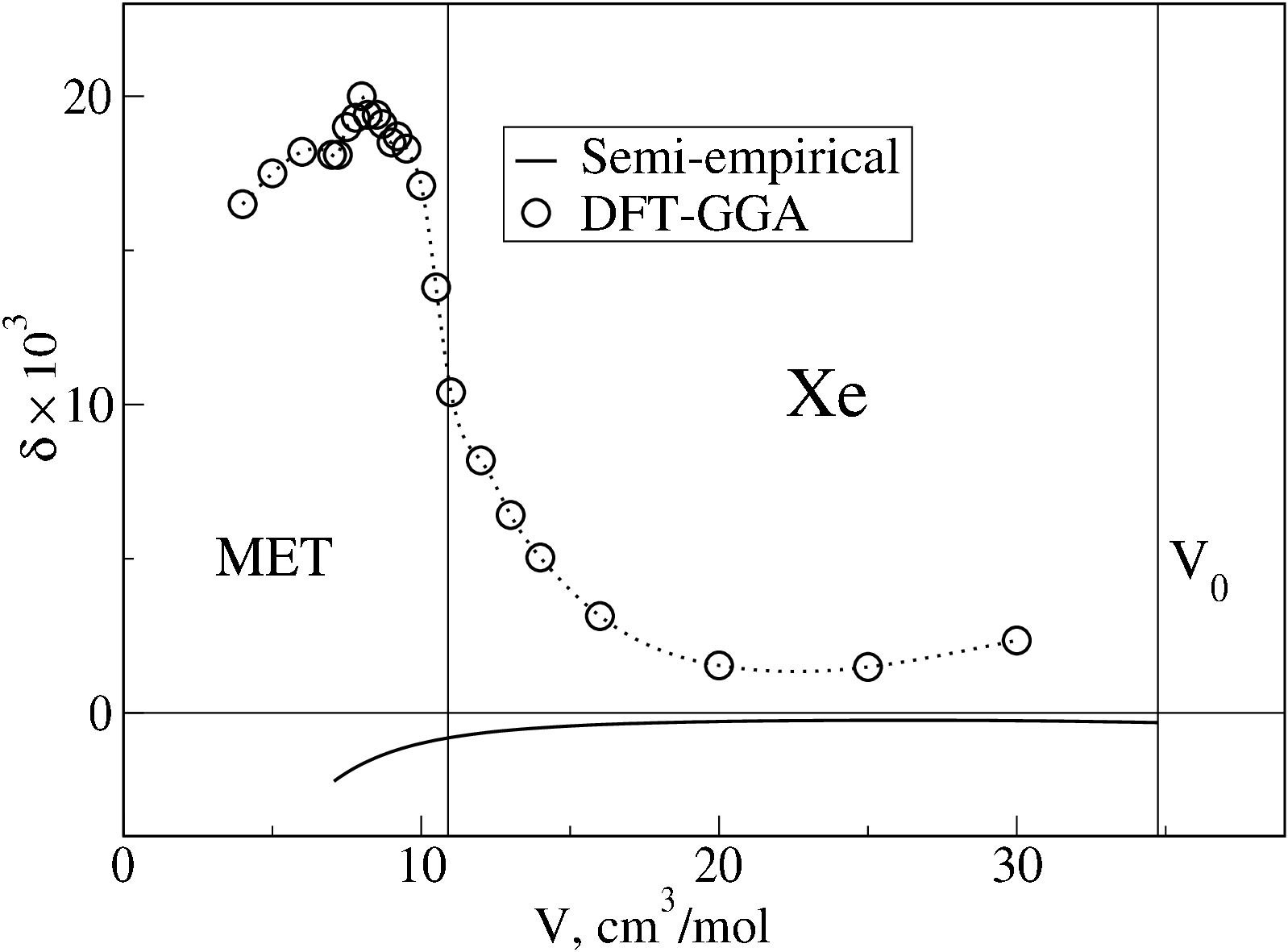}
\caption{\label{f:xe}The lattice distortion parameter $\delta$ for hcp xenon as a function of volume $V$ from
DFT-GGA and the semi-empirical (SE) potentials. The circles are calculated DFT-GGA points, while the
smooth dotted curve is a cubic spline.
}
\end{figure}

The calculated lattice distortion parameter $\delta$ as a function of
the molar volume $V$ for hcp helium is shown in 
Fig. \ref{f:he}. This is an extension of our previous work on He \cite{freiman09prb80:094112} with more volume
points included in the DFT-GGA calculation.
First of all, the order of magnitude of $\delta$ is about $10^{-3}$ in a wide range of
volumes (above 5 cm$^3$/mol).
This would be a very small effect for hcp metals, but it is an expected order of magnitude 
for molecular crystals.

The semi-empirical (SE) calculations without zero-point vibrations (dash-dotted curve) give  
negative $\delta$ in the entire volume range, with a maximum at about 8 cm$^3$/mol. The negative
sign is not unexpected, as most hcp solids have negative $\delta$, at least at low pressures.
At smaller volumes, negative $\delta$ grows in absolute value with decreasing volume 
(increasing pressure). It also grows sharply in absolute value when the volume
approaches 14 cm$^3$/mol. Note that the large equilibrium volume of solid helium (21.0 cm$^3$/mol) and the
preference of the hcp phase to fcc are effects of the zero-point vibrations (ZPV). 
If ZPV were not taken into account, 
helium at ambient conditions would be a solid with the fcc structure and 
molar volume of about 10 cm$^3$/mol \cite{kittel:solid_state_physics}. Therefore,
when comparing helium to other rare gas solids
(see the results below), the effective $V_0$ for He (without ZPV) should be placed at about
10 cm$^3$/mol, but not 21.0 cm$^3$/mol. Calculations without ZPV at $V>10$ cm$^3$/mol
correspond to the hypothetical case of expanded lattice and energetically unfavourable hcp structure, 
so there is little surprise that the $c/a$ ratio deviates from the ideal value at these volumes.

The picture changes drastically when the ZPV are included in the calculation
(solid line in Fig. \ref{f:he}). At $V<5$ cm$^3$/mol
the two curves practically coincide. However, the results with ZPV has no increase of 
the absolute value of $\delta$ at $V>10$ cm$^3$/mol.
In fact, there is no maximum and $\delta$ grows monotonously with increasing $V$. It stays
small in magnitude and possibly becomes positive at larger volumes. However, our calculations only include
ZPV in a rather simple way based on the Einstein model. For this reason, we do not present the results for volumes
larger than $14$ cm$^3$/mol. The detailed examination of the quantum crystal region is beyond the scope
of the present paper. In our previous paper \cite{freiman09prb80:094112} we have studied the effect
of pair (2-body) and triple (3-body) forces on lattice distortion. It has been shown that the triple forces
become important at approximately $V < 10$ cm$^3$/mol. It is difficult to say exactly at what volumes
the higher order terms (not included in the present SE calculations) come into play, but one can expect the
SE description to be adequate between at least 10 and 14 cm$^3$/mol (quite possibly down to 5 cm$^3$/mol).

The DFT calculations (circles and the dotted curve in Fig. \ref{f:he}) report a negative $\delta$ at 
$V < 5$ cm$^3$/mol, which grows in magnitude with decreasing $V$ much quicker than is given by the
SE potentials. The reason is that DFT automatically includes all n-body terms, while the SE method is limited
by the 3-body terms and should not be used for $V < 5$ cm$^3$/mol. For volumes above 5 cm$^3$/mol, however,
the $\delta$ from DFT behaves nonmonotonously and finally becomes positive and grows sharply at
$V > 11$ cm$^3$/mol. We believe the latter region (and possibly even the entire region above 5 cm$^3$/mol)
to be nonphysical, as GGA is known to describe poorly the van der Waals (vdW) interaction. However, for
$V < 4$ cm$^3$/mol the DFT is expected to be reliable, as the atomic repulsive forces dominate in this
high-pressure region and the vdW contribution is small. At very low volumes (which correspond to the
experimentally unreachable pressures of thousands of GPa), $\delta$ reaches a minimal value
of $-0.05$ at about 0.23 cm$^3$/mol, which is very close to the metallization point
(vertical line in the inset of Fig. \ref{f:he}). In the metallic phase,
$\delta$ increases sharply (thus decreasing in the absolute value) with decreasing $V$.

To summarize, DFT-GGA works for $V < 4$ cm$^3$/mol, while the SE potentials are expected
to give adequate results between 10 and 14 cm$^3$/mol. Unfortunately, one does not know what happens
exactly between 4 and 10 cm$^3$/mol, as DFT and SE give somewhat different results for this region.
However, both methods provide a negative (or close to zero) $delta$ of the order of 10$^{-4}$ - 10$^{-3}$
for these "intermediate" volumes. So now it is up for the experimentalists to clarify this issue.
It might be also tempting to apply one of the formally exact
quantum mechanical methods, such as quantum Monte Carlo or
configuration interaction, to this problem. However, these methods work with finite cluster, and as the
lattice distortion is a very delicate effect (energy differences can be as small as $10^{-9}$ Ry),
it would be very difficult to eliminate the finite cluster size errors and obtain a reliable result.
For this reason we do not use these methods in the present paper.
\subsection{Heavier RGS's}
The $\delta(V)$ dependence for hcp argon is presented in Fig. \ref{f:ar}.
The two vertical lines correspond to the metallization point and the equilibrium volume, respectively
(see Table \ref{t:v0m}).
The SE results (solid curve) are qualitatively similar
to the SE results for He (without ZPV). The effect of zero points vibrations in Ar and other heavier RGS's is very
small and the SE curve without ZPV (not shown) practically coincides with the full SE curve. 
The lattice distortion parameter $\delta$ is negative
with a maximum at about 15 cm$^3$/mol. The DFT-GGA results (circles and the dotted curve in fig. \ref{f:ar}),
however, are somewhat surprising. The GGA gives positive $\delta$ with a minimum around 10 cm$^3$/mol and $\delta$
increases quickly with decreasing $V$ below this point. It can be estimated that SE works above 16 cm$^3$/mol
and DFT-GGA works below 8 cm$^3$/mol. Apparently, neither method is adequate in the 
intermediate region between 8 and 16 cm$^3$/mol. One can only speculate that $\delta$
behaves more or less monotonously, crossing over from negative values of the order
of $3\times10^{-4}$ at higher volumes to the
positive values at lower volumes, as shown schematically by the dashed line in Fig. \ref{f:ar}.

The results for Kr and Xe are shown in Figs. \ref{f:kr} and \ref{f:xe} respectively.
The results are qualitatively very
similar to the case of Ar. Note that our DFT calculations clearly indicate positive $\delta$ at high pressures
for Xe, in good agreement with the experiment \cite{caldwell97sci277:930,jephcoat87prl59:2670} and the
previous calculations \cite{caldwell97sci277:930,yao07prb75:134104}, and
in contradiction to Ref. \cite{cohen97prb56:8575}. At $V=8$ cm$^3$/mol ($P \approx 300$ GPa),
$\delta$ for Xe has a sharp maximum. It could be related to
electronic topological transitions, but
the detailed analysis of the electronic structure
of the metallic phases is beyond the scope of the present paper. 

It is difficult to give a simple qualitative explanation to
$\delta$ being negative for He under pressure, and positive
for Ar, Kr and Xe. Since helium has two $1s$ electrons, while Ar, Kr and Xe have $np^6$ outermost shells,
one can assume that this physical property is mainly determined by the quantum number $l$ of
the outermost electron shell ($s$ or $p$) (the principal quantum number $n$ seems irrelevant).
It seems that the filled $p$ shell favors positive $\delta$ under pressure, while
the filled $1s$ shell of helium favors negative $\delta$.
Another possible explanation is that
the heavier RGS's undergo metallization at much smaller pressures compared to helium.
Note that the negative $\delta$ in helium rapidly decreases in absolute value in
the metallic phase below 0.23 cm$^3$/mol.
So if we assume that the closeness to the 
metallization favours positive $\delta$, it would explain the positive sign of
$\delta$ in Ar, Kr and Xe in the metallic phase and near the metallization,
while in He the effect is too weak (with the metalization point at 0.228 cm$^3$/mol).
In the language
of $n$-body interactions, the pair forces are described well with the Aziz potential (with
different parameters for different RGS's) and the 3-body potentials are likewise similar for
He, Ar, Kr and Xe. The triple forces always favour negative $\delta$, while
the pair forces give negative contribution to $\delta$ at larger volumes and positive
at smaller volumes \cite{freiman09prb80:094112}. The total effect of pair and triple forces
is always negative.  Apparently, the
distinction between He and the heavier RGS's manifests itself in 4-body and higher $n$-body terms,
which are missing in our semi-empirical calculations; while GGA automatically includes all $n$-body terms.

\section{\label{sec:conclusion} Conclusion}
We have calculated the lattice distortion parameter $\delta \equiv c/a - \sqrt{8/3}$
for the hcp rare gas solids He, Ar, Kr and Xe. At low pressures (covered by the semi-empirical
pair and 3-body potentials), $\delta$ is negative for all systems. At higher pressures (covered by
DFT-GGA), $\delta$ becomes positive for Ar, Kr and Xe, while it stays negative for helium.

\acknowledgments

The authors thank Dr. G. E. Grechnev for fruitful discussions. The work has been
supported by a scholarship of the President of Ukraine.

\bibliography{strings,dft,rare}

\end{document}